\documentclass[aps,prl,reprint,groupedaddress,showpacs,floatfix]{revtex4-1}

\usepackage{graphicx}
\usepackage{dcolumn}
\usepackage{bm}
\usepackage[mathlines]{lineno}

\begin{document}

\title{Unusual Electronic Structure of Few-Layer Grey Arsenic:
       A Computational Study}

\author{Zhen Zhu}
\affiliation{Physics and Astronomy Department,
             Michigan State University,
             East Lansing, Michigan 48824, USA}

\author{Jie Guan}
\affiliation{Physics and Astronomy Department,
             Michigan State University,
             East Lansing, Michigan 48824, USA}

\author{David Tom\'{a}nek}
\email%
{tomanek@pa.msu.edu}%
\affiliation{Physics and Astronomy Department,
             Michigan State University,
             East Lansing, Michigan 48824, USA}

\date{\today} 

\begin{abstract}
We use {\em ab initio} density functional theory to study the
equilibrium geometry and electronic structure of few-layer grey
arsenic. In contrast to the bulk structure that is semimetallic,
few-layer grey As displays a significant band gap that depends
sensitively on the number of layers, in-layer strain, layer
stacking and inter-layer spacing. A metal-semiconductor transition
can be introduced by changing the number of layers or the in-layer
strain. We interpret this transition by an abrupt change in the
spatial distribution of electronic states near the top of the
valence band.
%
\end{abstract}




\maketitle



There is growing interest in two-dimensional semiconductors with a
significant fundamental band gap and a high carrier mobility.
Whereas obtaining a reproducible and robust band gap has turned
into an unsurmountable obstacle for
graphene~\cite{{PKimPRL07},{graphane}}, the presence of heavy
transition metal atoms in layered dichalcogenide compounds limits
their carrier mobility\cite{Fuhrer2013}. Few-layer structures of
layered phosphorus allotropes, such as black phosphorus, are
rapidly attracting attention due to their combination of high
mobility and significant band
gaps~\cite{{DT229},{DT230},{Li2014}}. We find it conceivable that
other isoelectronic systems, such as arsenic, may display similar
structural and electronic properties as few-layer phosphorene
while being chemically much less reactive~\cite{AsStable}. In this
respect, the most abundant grey arsenic allotrope is the
structural counterpart of the layered A7 or blue
phosphorus~\cite{DT230}. Arsenic is commonly known for its
toxicity, which is highest for the yellow As allotrope and should
not be of concern for few-layer nanostructures. Whereas
crystalline grey arsenic displays rhombohedral stacking of layers
and is semimetallic~\cite{%
{Bullett1975},{Madelung},{Xu1993},{Tokailin1984},{Gonze1990},%
{Golin1965a},{Golin1965b}}, loss of crystallinity opens a
fundamental band gap in the amorphous
structure~\cite{{Bullett1976},{Madelung}}. Even though few-layer
grey arsenic has not been studied yet, analogies with blue
phosphorene make few-layer arsenic a plausible candidate for a 2D
semiconductor.


Here we use {\em ab initio} density functional theory (DFT) to
study the equilibrium geometry and electronic structure of
few-layer grey arsenic. This allotrope closely resembles the
structure of rhombohedral graphite with the exception of puckering
within the honeycomb lattice of the layers. In contrast to the
bulk structure that is semimetallic, few-layer grey As displays a
substantial band gap that depends sensitively on the number of
layers, in-layer strain, layer stacking and inter-layer spacing. A
metal-semiconductor transition can be introduced by changing the
number of layers or the in-layer strain. We interpret this
transition by an abrupt change in the spatial distribution of
electronic states near the top of the valence band.


Our computational approach to gain insight into the equilibrium
structure, stability and electronic properties of arsenic
structures is based on {\em ab initio} density functional theory
as implemented in the \textsc{SIESTA}~\cite{SIESTA} and
VASP\cite{VASP} codes. We use periodic boundary conditions
throughout the study, with multilayer structures represented by a
periodic array of slabs separated by a vacuum region
${\agt}15$~{\AA}. Unless specified otherwise, we use the
Perdew-Burke-Ernzerhof (PBE)~\cite{PBE} exchange-correlation
functional for most calculations. Selected results are compared to
the Local Density Approximation
(LDA)~\cite{{Ceperley1980},{Perdew81}} and other functionals
including the {\textsc optB86b-vdW}
functional~\cite{{Klimes10},{Klimes11}} that provides a better
description of van der Waals interactions and the {\textsc
HSE06}\cite{{HSE03},{HSE06}} hybrid functional. In our
\textsc{SIESTA} calculations we use norm-conserving
Troullier-Martins pseudopotentials~\cite{Troullier91}, and a
double-$\zeta$ basis including polarization orbitals. The
reciprocal space is sampled by a fine grid~\cite{Monkhorst-Pack76}
of $16{\times}16{\times}1$~$k$-points in the Brillouin zone of the
primitive unit cell for 2D structures and
$16{\times}16{\times}3$~$k$-points for the bulk. We use a mesh
cutoff energy of $180$~Ry to determine the self-consistent charge
density, which provides us with a precision in total energy of
${\alt}2$~meV/atom. All geometries have been optimized using the
conjugate gradient method~\cite{CGmethod}, until none of the
residual Hellmann-Feynman forces exceeded $10^{-2}$~eV/{\AA}.
Equilibrium structures and energies based on \textsc{SIESTA} have
been checked against values based on the {\textsc VASP} code.

\begin{figure}[t]
\includegraphics[width=1.0\columnwidth]{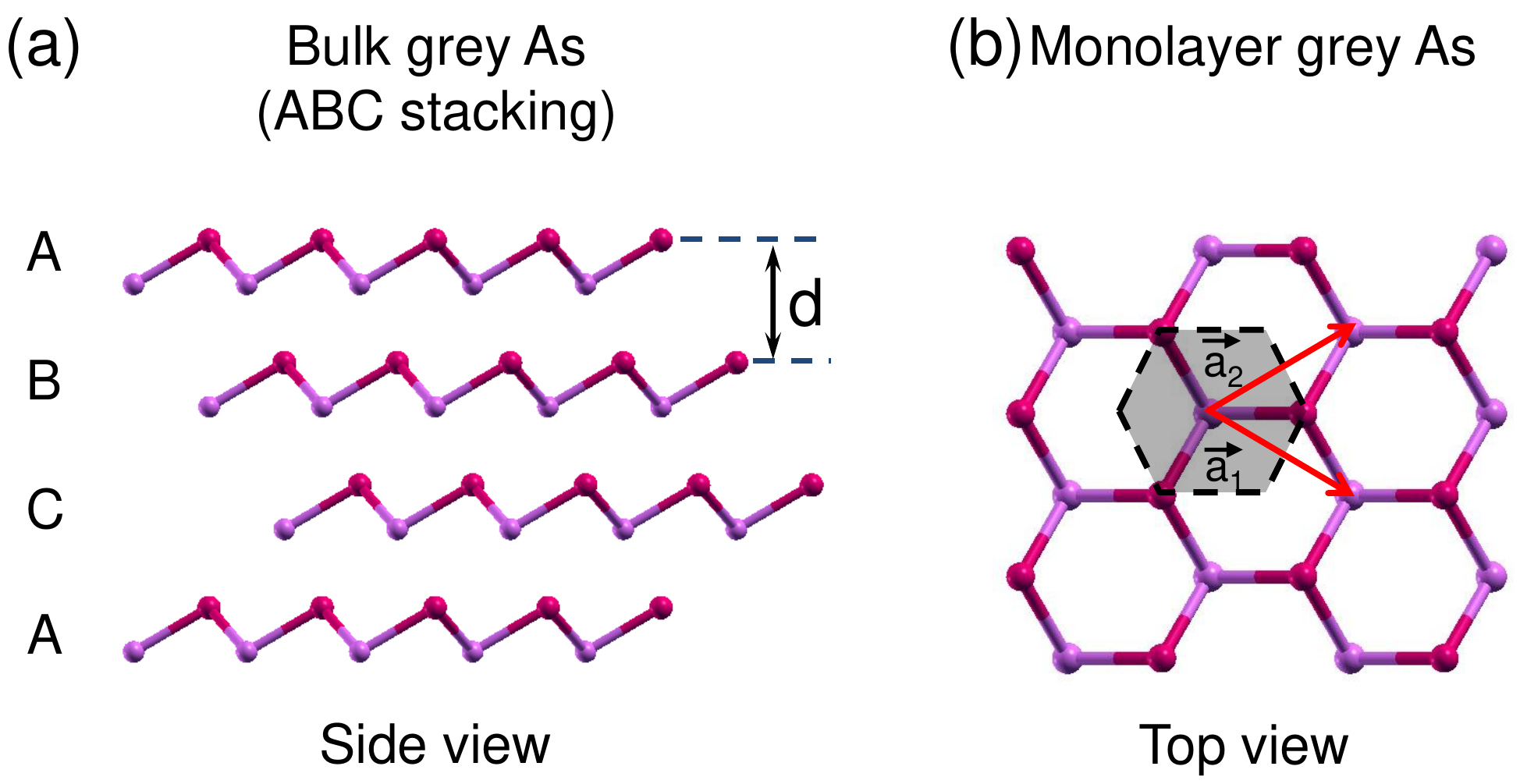}
\caption{(Color online) (a) Side view of the rhombohedrally (ABC)
stacked layered structure of bulk grey arsenic. (b) Top view of
the puckered honeycomb structure of a grey arsenic monolayer.
Atoms at the top and bottom of the non-planar layers are
distinguished by color and shading and the Wigner-Seitz cell is
shown by the shaded region. \label{fig1}}
\end{figure}

\begin{figure*}[t]
\includegraphics[width=2.0\columnwidth]{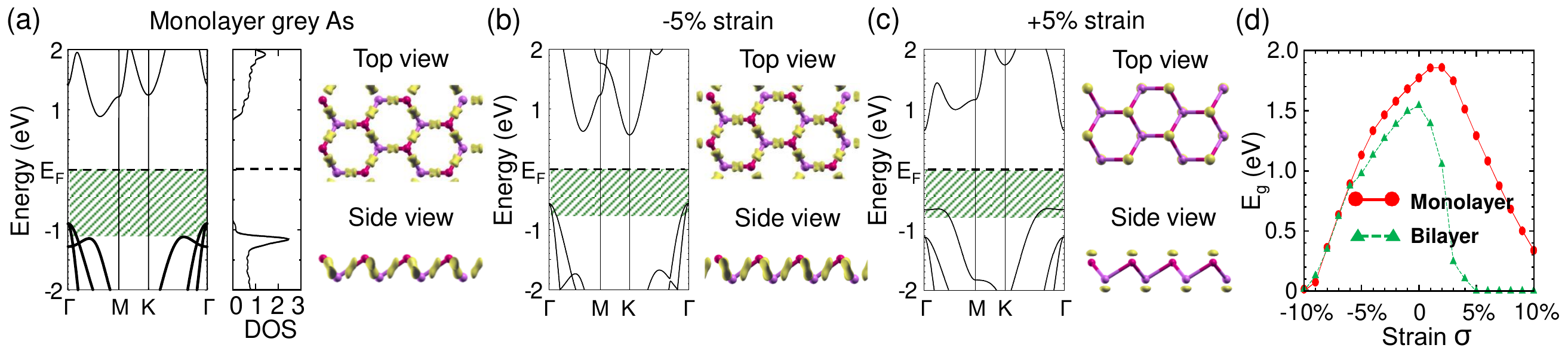}
\caption{(Color online) Electronic structure of (a) a relaxed
monolayer of grey As, and the same monolayer subject to a uniform
in-layer strain of (b) -5\% and (c) +5\%. The energy range between
the Fermi level $E_F$ and 0.2~eV below the top of the valence band
is green shaded in the band structure in the left panels. Electron
density $\rho_{vb}$ associated with these states, superposed with
a ball-and-stick model of the structure, is shown in the right
panels. (d) Dependence of the fundamental band gap $E_g$ on the
in-layer strain $\sigma$ in a monolayer and an AB-stacked bilayer
of grey As. \label{fig2}}
\end{figure*}

\begin{figure*}[t]
\includegraphics[width=2.0\columnwidth]{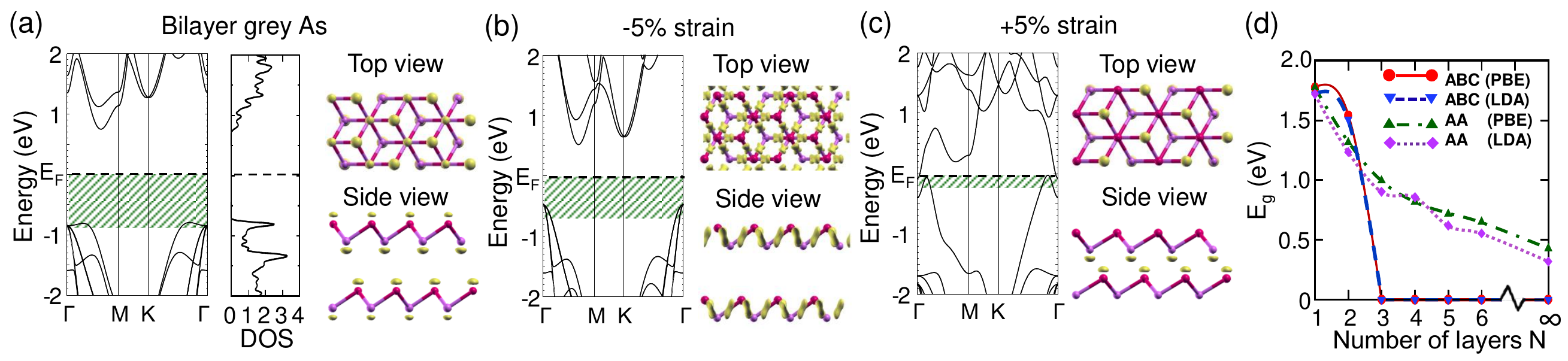}
\caption{(Color online) Electronic structure of (a) a relaxed
AB-stacked bilayer of grey As, and the same bilayer subject to a
uniform in-layer strain of (b) -5\% and (c) +5\%. The energy range
between the Fermi level $E_F$ and 0.2~eV below the top of the
valence band is green shaded in the band structure in the left
panels. Electron density $\rho_{vb}$ associated with these states,
superposed with a ball-and-stick model of the structure, is shown
in the right panels. (d) Dependence of the fundamental band gap
$E_g$ on the number of layers $N$ and the DFT functional in
few-layer As with ABC and AA stacking. The lines in (d) are guides
to the eye. \label{fig3}}
\end{figure*}


In contrast to the AB-stacked isoelectronic black phosphorus, bulk
grey arsenic prefers the rhombohedral (or ABC) layer stacking,
with the optimized structure shown in Fig.~\ref{fig1}. The
monolayer of grey As, depicted in top view in Fig.~\ref{fig1}(b),
resembles the honeycomb lattice of graphene with two atoms per
unit cell. Unlike planar graphene, however, the unit cell is
puckered, similar to blue (or $A7$) phosphorus~\cite{DT230}, and
very different from layered black phosphorus~\cite{DT229}.
Interatomic interactions within a monolayer are covalent,
resulting in a nearest-neighbor distance of 2.53~{\AA}. The phonon
spectrum of a grey arsenic monolayer, displayed in the
Supplemental Material~\cite{SM-2DAs14}, shows no soft modes, thus
indicating the stability of a free-standing monolayer.

\begin{table}[b]
\caption{Observed and calculated properties of layered grey
arsenic. $a=|\vec{a_1}|=|\vec{a_2}|$ is the in-plane lattice
constant and $d$ is the interlayer separation, as defined in
Fig.~\protect\ref{fig1}. $E_{coh}$ is the cohesive energy and
$E_{il}$ is the interlayer interaction energy per atom.}
\begin{tabular}{lcccc}
\hline %
Structure   & Bulk(ABC) %
            & Bulk(ABC)  & Bulk(AA) & Monolayer \\%
            & (expt.)    & (theory)   %
            & (theory)   & (theory) \\%
\hline %
$a$~({\AA}) & 3.76%
\protect\footnote{Experimental data of Ref.~\cite{Schiferl1969}.}
                           & 3.85\protect\footnote{Results based on the DFT-PBE functional~\cite{PBE}.}%
                           & 3.65$^{\rm b}$         &   3.64$^{\rm b}$ \\%
                           & --
                           & 3.85\protect\footnote{Results based on the LDA~\cite{Ceperley1980}.}
                           & 3.64$^{\rm c}$         &   3.61$^{\rm c}$ \\%
                           & --
                           & 3.82\protect\footnote{Results based on the optB86b van der
                                                   Waals functional~\cite{{Klimes10},{Klimes11}}.}
                           & 3.62$^{\rm d}$         &   3.58$^{\rm d}$ \\%
\hline %
$d$~({\AA})
                           & 3.52$^{\rm a}$         &   3.58$^{\rm b}$       %
                           & 5.15$^{\rm b}$         &   -- \\%
                           & --    & 3.46$^{\rm c}$  & 4.20$^{\rm c}$         &   -- \\%
                           & --    & 3.47$^{\rm d}$ & 4.31$^{\rm d}$         &   -- \\%
\hline %
$E_{coh}$                  & 2.96%
\protect\footnote{Experimental data of Ref.~\cite{Kittel}.}
           &   2.86$^{\rm b}$       %
                           & 2.85$^{\rm b}$         &   2.84$^{\rm b}$  \\%
(eV/atom)                  & --           &   3.60$^{\rm c}$       %
                           & 3.53$^{\rm c}$         &   3.45$^{\rm c}$ \\%
\hline %
$E_{il}$                   & --           &   0.02$^{\rm b}$       %
                           & 0.01$^{\rm b}$         &   -- \\%
(eV/atom)                  & --           &   0.16$^{\rm c}$       %
                           & 0.10$^{\rm c}$         &   -- \\%
& --           &   0.17$^{\rm d}$       %
               & 0.13$^{\rm d}$         &   -- \\%

\hline
\end{tabular}
\label{table1}
\end{table}

Observed and calculated structural and cohesive properties in grey
arsenic are summarized in Table~\ref{table1}. The calculated
inter-layer separation in the ABC-stacked bulk system is
$d=3.58$~{\AA}, similar to the observed value~\cite{Schiferl1969}.
The interlayer interaction energy of ${\approx}0.02$~eV/atom,
based on PBE, is slightly higher than in blue
phosphorus~\cite{DT230}. While this value is likely
underestimated, the optB86b value of $0.17$~eV/atom and the LDA
value of $0.16$~eV/atom likely overestimate the interlayer
interaction, as discussed in the Supplemental
Material~\cite{SM-2DAs14}. The low interlayer interaction energy,
similar to graphite and black phosphorus, suggests that few-layer
As may be obtained by mechanical exfoliation from the bulk
structure. The small difference in the length of the in-plane
lattice vectors $a=|\vec{a}_1|=|\vec{a}_2|=3.64$~{\AA} in the
isolated monolayer and $a=3.85$~{\AA} in the bulk structure is
also consistent with a weak interlayer interaction. In agreement
with the experiment, we find AA-stacked grey arsenic to be less
stable than the ABC-stacked structure, even though the energy
difference lies within 10~meV/atom. The optimum inter-layer
separation in the less favorable AA stacking increases to
$d=5.15$~{\AA}.


In agreement with observations~\cite{Madelung}, our DFT results
indicate that bulk grey arsenic is semimetallic. Our corresponding
DFT results for the electronic structure of a monolayer of grey
arsenic are presented in Fig.~\ref{fig2}. In stark contrast to the
bulk, the monolayer structure is semiconducting with an indirect
fundamental band gap $E_{g}{\approx}1.71$~eV. Comparison with more
precise {\textsc HSE06}~\cite{{HSE03},{HSE06}} hybrid functional
calculations, discussed in the Supplemental
Material~\cite{SM-2DAs14}, indicates that the PBE value of the
band gap is likely underestimated by ${\agt}0.4$~eV in few-layer
grey arsenic as a common shortcoming of DFT. Still, the electronic
structure of the valence and the conduction band region in DFT is
believed to closely represent experimental results. Therefore, we
expect the charge density associated with frontier orbitals near
the top of the valence band, shown in Fig.~\ref{fig2}(a), to be
represented accurately. These states correspond to the energy
range highlighted by the green shading in the band structure
plots, which extends from mid-gap to 0.2~eV below the top of the
valence band. As seen in the right panel of Fig.~\ref{fig2}(a),
these frontier orbitals lie in the region of the interatomic bonds
and are dominated by in-plane $p$-orbitals.

As seen in the $E(\vec{k})$ plot in Fig.~\ref{fig2}(a), states
near the top of the valence band at $\Gamma$ display a strong
dispersion. This is a signature of a very low hole mass, caused by
a well connected network of frontier orbitals, which are depicted
in the right panel of Fig.~\ref{fig2}(a). This is quite different
from black phosphorene, where the frontier valence band orbitals
are dominated by out-of-plane $p$-orbitals with little overlap,
which reduces the band dispersion and thus increases the hole
mass. We find that the effective mass near $\Gamma$ in few-layer
arsenic is not only lower, but -- in contrast to black
phosphorene~\cite{{Yang2014},{DT229}} -- also isotropic. Since
high carrier mobility values have been reported in bulk grey
arsenic~\cite{Madelung}, we believe that also few-layer arsenic
may display a higher mobility than few-layer black phosphorus.

As seen in Fig.~\ref{fig2}(b), a uniform 5\% in-layer compression
reduces the band gap, but keeps it indirect and does not change
drastically the character of the frontier orbitals. According to
Fig.~\ref{fig2}(d), compression in excess of 10\% would close the
band gap, turning the monolayer metallic. Interestingly, also a
uniform in-layer stretch reduces the band gap in the monolayer
significantly. As seen in Fig.~\ref{fig2}(c), stretching the
monolayer by 5\% moves the bottom of the conduction band near
$\Gamma$. The valence band with an energy eigenvalue $-1.3$~eV at
$\Gamma$ in the relaxed monolayer, shown in Fig.~\ref{fig2}(a),
moves up and becomes the top valence band in the stretched layer
in Fig.~\ref{fig2}(c). This band gradually flattens near $\Gamma$
upon stretching, and the monolayer becomes a direct-gap
semiconductor at $\sigma{\agt}+6$\%.

More important is the change in the character of the frontier
orbitals caused by the motion of this band relative to the other
valence bands near $\Gamma$. In contrast to the relaxed and
compressed layer depicted in Figs.~\ref{fig2}(a) and
~\ref{fig2}(b), the frontier orbitals in the stretched layer are
dominated by $p$-orbitals normal to the layer. As we will see
later, the character change of the frontier orbitals from in-plane
to out-of-the-plane has an important effect on the electronic
structure of multi-layer systems near $E_F$. In the simplest
example of this difference, shown in Fig.~\ref{fig2}(d), which
compares a bilayer to a monolayer, the band gap value is
essentially the same in both systems during compression, since the
frontier orbitals in the layers remain in the plane and do not
overlap. The character change of the frontier orbitals to
out-of-plane upon stretching is the main cause for the large
difference in the band gap value between a monolayer and a
bilayer. Quite significant in this respect is our finding that a
bilayer should turn metallic for $\sigma{\agt}5$\%.

A closer look at the behavior of frontier orbitals in a bilayer
subject to different levels of strain is offered in
Fig.~\ref{fig3}. As seen in Fig.~\ref{fig3}(a), the top valence
band in the relaxed bilayer has a similar dispersion in the
electronic structure near $\Gamma$ as a stretched monolayer in
Fig.~\ref{fig2}(c). Also the right panels of the two sub-figures
confirm the similar character of the frontier orbitals in a
stretched monolayer and a relaxed bilayer. Since the bottom of the
conduction band in the relaxed bilayer is not at $\Gamma$, this
system is an indirect-gap semiconductor with a narrower gap than
the relaxed monolayer.

Upon compression, the band ordering near the top of the valence
band changes in the bilayer. As seen in Fig.~\ref{fig3}(b), the
character of the frontier orbitals in a system subject to 5\%
uniform compression changes to in-plane, similar to a compressed
monolayer. A compressed arsenic bilayer remains an indirect-gap
semiconductor.

As seen in the right panel of Fig.~\ref{fig3}(c), a uniform
$\sigma=+5$\% strain in the bilayer changes the character of the
frontier orbitals from in-plane to out-of-plane, similar to our
findings for the stretched monolayer reported in
Fig.~\ref{fig2}(c). The frontier orbitals, which are primarily
distributed in the inter-layer region, bring the two layers closer
together. The increased hybridization at a decreased interlayer
distance accelerates the band gap closure and, for
$\sigma{\agt}+5$\%, turns the bilayer to a semi-metal.

The overall dependence of the fundamental band gap on the number
of layers $N$ and the stacking sequence, depicted in
Fig.~\ref{fig3}(d), shows a uniform trend of band gap reduction
with increasing value of $N$, which has been noted also for the
different layered allotropes of phosphorus~\cite{{DT229},{DT230}}.
In particular, we find ABC-stacked arsenic slabs with $N>2$ to be
metallic. We also observe notable changes in the optimum
inter-layer distance $d$ with changing number of layers and
stacking sequence, which strongly affect the electronic structure.
We find the optimum inter-layer distance in AA-stacked structures
to be much larger than in ABC-stacked structures, which slows down
the reduction of the band gap with growing $N$. To check the
validity of this trend, we reproduced the band gap values obtained
using both PBE and LDA exchange-correlation functionals for
structures optimized by PBE in Fig.~\ref{fig3}(d). Small
deviations from this trend, associated with the specific
functionals, are discussed in the Supplemental
Material~\cite{SM-2DAs14}.

The weak inter-layer interaction in layered grey arsenic should
allow for a mechanical exfoliation of few layer structures in
analogy to graphene and phosphorene. Besides mechanical
exfoliation, few-layer grey arsenic has been successfully
synthesized by Molecular Beam Epitaxy (MBE), the standard method
to grow GaAs, by supplying only As in the growth
chamber~\cite{arsenic1993}. Chemical Vapor Deposition (CVD), which
had been used successfully in the past to grow
graphene\cite{{KimNat2009},{ReinaNL2009}} and
silicene\cite{VogtPRL2012}, may become ultimately the most common
approach to grow few-layer grey arsenic on specific substrates.
Substrates such as Ag(111), or even Zr(0001) and Hf(0001) should
be advantageous to minimize the lattice mismatch during CVD
growth.

From the viewpoint of electronic applications, an ideal 2D
semiconductor should combine a sizeable fundamental band gap with
a high carrier mobility and chemical stability. Equally important
is identifying a suitable way to make electrically transparent
contacts. Graphene, with the exception of its vanishing band gap,
satisfies the latter three criteria ideally. Transition metal
dichalcogenides, including MoS$_2$, bring the benefit of a nonzero
band gap, but display lower intrinsic carrier mobility due to
enhanced electron-phonon coupling, primarily caused by the
presence of heavy elements such as Mo, and suffer from high
tunneling barriers at contacts to the chalcogen atoms. Few-layer
systems of black phosphorus and arsenic do show a sizeable band
gap, exhibit a higher carrier mobility than MoS$_2$~\cite{DT229}
and the ability to form transparent contacts to metal leads. Of
the two group V elements, the heavier arsenic appears more
resilient to oxidation~\cite{AsStable}. If indeed few-layer grey
arsenic turns out to be chemically stable, it may become an
excellent contender for a new generation of nano-electronic
devices.


In conclusion, we have used {\em ab initio} density functional
theory to study the equilibrium geometry and electronic structure
of few-layer grey arsenic. This allotrope closely resembles the
structure of rhombohedral graphite with the exception of puckering
within the honeycomb lattice of the layers. In contrast to the
bulk structure that is semimetallic, we found that few-layer grey
As displays a significant band gap that depends sensitively on the
number of layers, in-layer strain, layer stacking and inter-layer
spacing. A metal-semiconductor transition can be introduced by
changing the number of layers or the in-layer strain. We showed a
relationship between this transition and an abrupt change in the
spatial distribution of electronic states near the top of the
valence band. Due to the weak inter-layer interaction, grey
arsenic should exfoliate easily to form few-layer structures.
Alternative ways to synthesize few-layer arsenic include MBE and
CVD.

\begin{acknowledgements}
We acknowledge useful discussions with Gotthard Seifert and Bilu
Liu. Computational resources for this study have been provided by
the Michigan State University High Performance Computing Center.
\end{acknowledgements}



%

\end{document}